\documentclass[11pt, singlecolumn, citestyle=authoryear]{article}
\usepackage{sectsty,titling}
\usepackage{tcolorbox}
\usepackage{authblk}
\usepackage{enumitem}
\usepackage{listings}
\usepackage{xcolor}
\usepackage[hidelinks]{hyperref}
\usepackage{multirow}
\usepackage[normalem]{ulem}

\usepackage{authblk}
\usepackage{academicons}
\usepackage{array}

\usepackage{longtable}
\usepackage[frozencache=true,cachedir=minted-cache]{minted} 

\definecolor{codegreen}{rgb}{0,0.6,0}
\definecolor{codegray}{rgb}{0.5,0.5,0.5}
\definecolor{codepurple}{rgb}{0.58,0,0.82}
\definecolor{backcolour}{rgb}{0.95,0.95,0.92}

\definecolor{customcolor}{RGB}{153,204,255}
\definecolor{astral}{RGB}{46,116,181}
\colorlet{coverlinecolor}{customcolor}

\subsectionfont{\color{astral}}
\sectionfont{\color{astral}}

\lstdefinestyle{mystyle}{
    backgroundcolor=\color{backcolour},   
    commentstyle=\color{codegreen},
    keywordstyle=\color{magenta},
    numberstyle=\tiny\color{codegray},
    stringstyle=\color{codepurple},
    basicstyle=\ttfamily\footnotesize,
    breakatwhitespace=false,         
    breaklines=true,                 
    captionpos=b,                    
    keepspaces=true,                 
    numbers=left,                    
    numbersep=5pt,                  
    showspaces=false,                
    showstringspaces=false,
    showtabs=false,                  
    tabsize=2
}
\tcbuselibrary{all}

\newtcolorbox{simpleBox}[2]{enhanced,attach boxed title to top center={yshift=-3mm,yshifttext=-1mm},colback=#2!5!white,colframe=#2!75!black,colbacktitle=#2!80!black,title={#1},fonttitle=\bfseries, boxed title style={size=small,colframe=#2!75!black}}

\newtcolorbox{fancyBox}[1]{enhanced,frame style image=blueshade.png,
  opacityback=0.75,opacitybacktitle=0.25,
  colback=blue!5!white,colframe=blue!75!black,
  title={#1}}

\setlist[itemize]{label=\color{astral}\textbullet}

\newcommand{\orcid}[1]{\href{https://orcid.org/#1}{\textcolor[HTML]{A6CE39}{\includegraphics{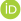}}}}

\begin{document}
\setcounter{Maxaffil}{0}
\title{\bfseries \sffamily \Huge \color{astral} BinCodex \\
    \bfseries \LARGE LISA BPS Common Output Format v. 1.1.0}

\author{Ruggero Valli\footnote{\url{ruvalli@mpa-garching.mpg.de}} \orcid{0000-0003-3456-3349}}
\affil{Max-Planck-Institut f{\"u}r Astrophysik, Karl-Schwarzschild-Straße 1, 85741 Garching, Germany}
\author{Luca Graziani\footnote{\url{luca.graziani@uniroma1.it}} \orcid{0000-0002-9231-1505}}
\affil{Dipartimento di Fisica, Sapienza, Universit{\`a} di Roma, Piazzale Aldo Moro 5, 00185, Roma, Italy}

\author[]{the LISA Synthetic UCB Catalogue Group}
\affil{Complete author list at the end of this document}

\maketitle

\section{Introduction}
This document describes BinCodex, a common format for the output of binary population synthesis (BPS) codes agreed upon by the members of the LISA Synthetic UCB Catalogue Group. The goal of the format is to provide a common reference framework to describe the evolution of a single, isolated binary system or a population of isolated binaries.

\subsection{Motivation}
To provide a common output format across BPS codes participating in the LISA collaboration to: 
\begin{itemize}
    \item Simplify data processing, especially for large datasets. 
    \item Easily and safely compare different BPS codes predictions.
    \item Minimize errors due to unit conversion, and avoid misinterpretation of similar physical quantities, by adopting a common definition.
    \item Provide a standardized framework of physical quantities facilitating the interpretation of BPS codes scientific outputs. 
\end{itemize}

\subsection{Leading Principles}
The proposed common format is designed to adhere to the following general principles:

\textbf{Progressive}: the common format is designed to be forward/backward-compatible with its extensions and additions. This means that the common format is aimed at being capable of changing in time to comply with the evolution of the BPS codes currently in the collaboration (and their future modifications), as well as to include new BPS codes.

\textbf{Flexible}: the common format can accommodate both specific and generic information, as well as missing data.

\textbf{Unambiguous}: the common format defines the value, organization and labelling of all required data and its compliance as precisely as possible to minimize differences in its practical implementation.

\textbf{Easy to implement}: the proposed format, in its initial version, aims to be close to the data formats currently adopted by existing BPS codes.

To satisfy the previous requirements the format adopts a hierarchical organization of its values in stratified tiers dubbed as T0 (common core variables), T1 and T2. Their definition is provided below and could change and vary across progressive versions of the common format. As a rule, a file containing data defined as compliant with BinCodex should be provided by specifying the reference format version and the implemented tier in the header.

\section{Description of the format}

BinCodex can be practically implemented in both \texttt{HDF5} and \texttt{ASCII} \texttt{CSV} file formats. In this version, we describe its \texttt{CSV} implementation.
An \texttt{ASCII} \texttt{CSV} file conforming to the present format is required to respect the following rules:

\begin{itemize}
    \item Lines starting with the character \texttt{'\#'} are called \textit{free-format} or comment lines. Their number is unspecified and their content does not follow any specific format rule.
    \item Lines starting with a character different form \texttt{'\#'} contain a sequence of fields separated by commas \texttt{','}.
    \item Quantities in double precision (i.e having a double value), are represented in exponential notation with a number of decimal digits $\ge 2$ (e.g. 1.23e10 or 1.23456e10). The standard suggests a maximum number of digits $< 5$.
    \item Quantities of type string (not present in a comment line)  can contain all printable \texttt{ASCII} characters, except \texttt{','} and \texttt{'\#'}, which are defined as reserved characters.
\end{itemize}

A \texttt{BinCodex}-compliant file is composed of three parts: \textbf{the Free-Format section} (\S \ref{sec:free-format}), \textbf{the Header} (\S \ref{sec:header}) and \textbf{the Table of Values} (\S \ref{sec:table}).

\subsection{The Free-Format section}
\label{sec:free-format}
The Free-Format section is intended to store simulation-specific parameters as well as comments and information regarding the run generating the stored dataset. The section is composed of comment lines only, as defined above. It ends at the first line of the file that does not start with \texttt{'\#'}. The user can freely choose both the content and the number of lines of a Free-Format section. 

\begin{simpleBox}{Examples}{astral}

    Some possible use cases for the free format section are listed below.
   \begin{itemize}
        \item Each simulation run is generally identified by a timestamp, a specific version of the adopted BPS code binary or even a build-release of the code. A simulation unique ID or name is also often adopted to uniquely label different runs. All the above information is likely stored in the Free-Format section of the file if the fields defined in the Header Section are not sufficient.
        \item When performing several runs for a parameter exploration study, the Free-Format section can be used to store the adopted ranges of values.
        \item The exact value of physical constants used in the code can be stored in the Free-Format section to guarantee consistency with post-processing operations requiring unit conversions.
    \end{itemize}

    NOTE: It is not advisable to write in the Free-Format section the values of \textit{all} the set-up parameters adopted in the run, because many of them are likely to be specific to a particular code and it would considerably clutter the header. In this sense the free-format section should not be considered a substitute for proper run documentation or an associated log file. The best way to make simulation results reproducible is to provide your run-scripts, the version of a code binary as well as the code documentation.
\end{simpleBox}

\subsection{The Header}
\label{sec:header}

The Header section is intended to provide a minimum set of parameters uniquely associating the stored data with the \texttt{BinCodex} version as well as the generating BPS code binary. It begins at the first line that does not start with \texttt{'\#'} and it is composed of 3 lines. The first 2 lines contain the same number of columns, each containing a field. The first line contains the unique labels of the fields, while the second line contains their specific values, appearing in the same order as the labels they correspond to. The required header fields are described in Table \ref{tab:header}.
The third line of the header contains the unique labels of the variables describing each binary system evolution according to the specific implemented tier (either defined in T0, T1 or T2, see table \ref{tab:columns}). Their physical corresponding values are listed in columns, in the same order as the Table of Values.

\begin{simpleBox}{Examples}{astral}
An example of a Header compliant with the standard is the following:
\begin{lstlisting}
cofVer,cofTier,cofExtension,bpsName,bpsVer,contact,NSYS,NLINES,Z
1.0,T0,,ComBinE,1.0.0,pippo@pluto.com,0.01,1000,12345
ID,UID,time,event,semiMajor,eccentricity,type1,mass1,radius1,Teff1,massHeCore1,type2,mass2,radius2,Teff2,massHeCore2
\end{lstlisting}
\end{simpleBox}

\begin{simpleBox}{Forward-compatibility: the Header}{red}
To ensure forward-compatibility, the only requirement is that \textbf{all mandatory labels described in Table \ref{tab:header} are present in the Header}. The order in which they appear is not fixed. Moreover, additional fields that are not described in Table \ref{tab:header} are ignored. In this way, future extensions of the format that add new header fields will be compatible with parsers that are compliant with older versions.
\end{simpleBox}

\begin{fancyBox}{Data integrity check}
The two header fields \texttt{NSYS} and \texttt{NLINES} can be used as a simple check of the integrity of the file. Ensuring that the number of systems stored in the file and that the number of lines in the file corresponds to the values reported in the header helps identify problems during the copy or the manipulation of the file.
\end{fancyBox}

\begin{table}[ht]
    \centering
    \makebox[\linewidth]{
    \begin{tabular}{|c|l|c|c|c|c|}
        \hline
        Label & Description & Mandatory? & Type & Example\\
        \hline
        \texttt{cofVer}       & Common Format Version                                  & Y & string & 1.0 \\
        \texttt{cofTier}     & Common Format Tier  (\S \ref{sec:tiers})             & Y & string & T0 \\
        \texttt{cofExtension} & Name of the format extension (\S \ref{sec:extensions}) & Y & string & myExtension \\
        \texttt{bpsName}      & BPS code used to generate the data                          & Y & string & SEVN \\
        \texttt{bpsVer}       & BPS code version                                            & Y & string & 1.0.0 \\
        \texttt{contact}      & Contact person or email                                & Y & string & pippo@pluto.com \\
        \texttt{NSYS}         & Number of binary systems in the file                   & Y & long    & 1000000 \\
        \texttt{NLINES}       & Total number of lines in the file            & Y & long    & 12345678 \\
        \texttt{Z}            & Initial absolute metallicity.                          & N & double & 1.00e-2 \\
        \hline
        %\texttt{Ggrav}   & gravitational constant (cgs)      & cm$^3$ g$^{-1}$ s$^{-2}$ & double & 6.67430e-8 \\
        %\texttt{clight}  & speed of light (cgs)              & cm/s                     & double & 2.99792458e10 \\
        %\texttt{Msun}    & mass of the Sun (cgs)             & g                        & double & 1.98847e33 \\
        %\texttt{Rsun}    & radius of the Sun (cgs)           & cm                       & double & 6.957e10 \\
        %\texttt{Lsun}    & luminosity of the Sun (cgs)       & erg/s                    & double & 3.828e33 \\
        %\texttt{Zsun}    & metallicity of the Sun            &                          & double & 0.0134 \\
        %\texttt{AU}      & astronomical unit (cgs)           & cm                       & double & 1.495978707e13 \\
        %\texttt{day}    & day (cgs)                          & s                        & double & 8.64e4 \\
        %\texttt{year}   & year (cgs)                         & s                        & double & 3.1536e7 \\
        %\hline
    \end{tabular}
    }
    \caption{Fields of the Header section. Note that different \texttt{bpsName} labels should indicate different codes or different code branches developed independently, otherwise, the label version should mark different code releases.}
    \label{tab:header}
\end{table}

\subsection{The Table of Values}
\label{sec:table}
The evolution of binary systems, as a collection of their physical states, is contained in the Table of Values. 
After the third line of the header, the file contains a table of values corresponding, in the same order, to the columns listed at line 3 of the header. Each row describes the state of a binary system at a given point in time, as a collection of properties of the binary and of its components at fixed time. The states pertaining to the same systems must appear in contiguous rows and are likely ordered chronologically. If two consecutive lines pertain to the same system, the value of the \texttt{time} column should increase with the line number, unless the time between two states goes below the resolution of the code. To guarantee that consecutive states are uniquely identified, this standard assumes that the line printing order also defines their logical priority. 
Table \ref{tab:columns} describes the columns required by the format and their organization in tiers. The columns must be all present in the file once the implemented tier is established, but some of the columns may be left empty if not provided by the code (see box \textit{Undefined and missing values}).
The format also allows comment lines between binary system lines, while comment lines between state lines are forbidden. Comment lines separating binaries could be used, for example, to specify a different value of the assumed binary metallicity. In this case, the label \texttt{Z} is missing in the header file, while repeated in a comment line before a new system lineset. It should be noted though, that for large datasets a more scalable solution should be adopted; see Section \ref{sec:system-specific parameters} for more details. 

\subsubsection{Tiers}
\label{sec:tiers}
The meaning of the tiers is as follows:
\begin{itemize}
    \item Tier 0 (T0) is referred to as Common Core Tier, indicating the minimum set of physical variables describing a binary system's physical status at a fixed time (i.e. a binary state). All the BPS codes participating in the collaboration share values of T0 so that their physical outcome can be safely compared once provided in the common format.
    \item Tier 1 (T1) contains all the variables of T0, plus several variables shared by the largest group of participating BPS codes. Not all codes are expected to provide T1 values so the comparison across codes is limited to T0. T1 also serves to stimulate the development of missing features in BPS codes missing these quantities. 
    \item Tier 2 (T2) groups all the quantities in T0 and T1, plus several physical values provided by some of the participating BPS codes without the requirement of a common intersection. For specific applications or analyses, they could be necessary.
\end{itemize}

\begin{fancyBox}{Undefined and missing values}
    \label{box:undefined_and_missing}
    Unless differently specified, when the value of a field is not defined, it is assigned the value \texttt{NaN}. When the value of a field is missing/unknown/not reported, the field is left empty (i.e. two contiguous comma separators).
    
    For instance, when a binary system has been unbound, its orbital parameters (\texttt{eccentricity}, \texttt{semiMajor}, etc.) are not defined anymore, and those fields will be assigned the value \texttt{NaN}. On the other hand, if the BPS code does not provide the eccentricity in the output, the component may or may not have a defined eccentricity, but we do not know its value. The field will then be left empty.
    
    Other examples of quantities that may be not defined are the properties (mass, radius...) of an object that does not exist anymore (type = -1).
\end{fancyBox}

\begin{table}
\begin{tabular}{|c|l|c|c|}
\hline
Label & Description & Unit & Variable Type\\
\hline
\multicolumn{4}{|c|}{T0} \\
\hline
\texttt{ID}            & progressive system ID   &                                  & long   \\
\texttt{UID}           & unique system ID        &                                  & string \\
\texttt{time}          & time                    & Myr                              & double \\
\texttt{event}         & event type$^a$          &                                  & string    \\
\texttt{semiMajor}     & semi-major axis         & $R_\odot$                        & double \\
\texttt{eccentricity}  & orbital eccentricity    &                                  & double \\
\texttt{type1}         & type of object 1$^b$    &                                  & string    \\
\texttt{mass1}         & mass 1                  & $M_\odot$                        & double \\
\texttt{radius1}       & radius 1                & $R_\odot$                        & double \\
\texttt{Teff1}         & effective temperature 1 & K                                & double \\
\texttt{massHeCore1}   & He core mass 1          & $M_\odot$                        & double \\
\texttt{type2}         & type of object 2$^b$    &                                  & string    \\
\texttt{mass2}         & mass 2                  & $M_\odot$                        & double \\
\texttt{radius2}       & radius 2                & $R_\odot$                        & double \\
\texttt{Teff2}         & effective temperature 2 & K                                & double \\
\texttt{massHeCore2}   & He core mass 2          & $M_\odot$                        & double \\
\hline
\multicolumn{4}{|c|}{T1} \\
\hline
\texttt{envBindEn1}     & envelope binding energy 1    & erg                           & double \\
\texttt{envBindEn2}     & envelope binding energy 2    & erg                           & double \\
\texttt{massCOCore1}   & CO core mass 1             & $M_\odot$                     & double \\
\texttt{massCOCore2}   & CO core mass 2             & $M_\odot$                     & double \\
\texttt{radiusRT1}     & radius of the Roche Lobe 1 & $R_\odot$                     & double \\
\texttt{radiusRT2}     & radius of the Roche Lobe 2 & $R_\odot$                     & double \\
\texttt{period}        & orbital period             & day                           & double \\
\texttt{luminosity1}    & bolometric luminosity 1     & $L_\odot$                     & double \\
\texttt{luminosity2}    & bolometric luminosity 2     & $L_\odot$                     & double \\
\hline
\multicolumn{4}{|c|}{T2} \\
\hline
\texttt{dMdt1}         & mass gained/lost by 1      & $M_\odot/yr$                  & double \\
\texttt{dMdt2}         & mass gained/lost by 2      & $M_\odot/yr$                  & double \\
\texttt{jOrb}          & orbital angular momentum   & $M_\odot$ $R_\odot$$^2$/day   & double \\
\texttt{spin1}         & abs. value of spin 1       & $M_\odot$ $R_\odot$$^2$/day   & double \\
\texttt{spin2}         & abs. value of spin 2       & $M_\odot$ $R_\odot$$^2$/day   & double \\
\texttt{omega1}        & angular velocity 1         & day$^{-1}$                    & double \\
\texttt{omega2}        & angular velocity 2         & day$^{-1}$                    & double \\
\texttt{Hsup}          & H surface mass fraction    &                               & double \\
\texttt{Hesup}         & He surface mass fraction   &                               & double \\
\texttt{Csup}          & C surface mass fraction    &                               & double \\
\texttt{Nsup}          & N surface mass fraction    &                               & double \\
\texttt{Osup}          & O surface mass fraction    &                               & double \\
\texttt{spectralType1} & spectral type 1            & & string\\
\texttt{spectralType2} & spectral type 2            & & string\\
\hline
\end{tabular}
\caption{The columns present in the table. $^a$\footnotesize{See Section \ref{sec:component_types}.} $^b$\footnotesize{See Section \ref{sec:events}.}}
\label{tab:columns}
\end{table}

\begin{fancyBox}{ID and UID}
    There are two kinds of IDs. The \textit{progressive system ID} (\texttt{ID}) is an integer number that is assigned progressively to the systems. The first system appearing in the file will have \texttt{ID} 1, the second will have \texttt{ID} 2 and so on until the last system with \texttt{ID} \texttt{NSYS}. The \texttt{ID} is an easy way to identify a system within the same file. For example, for fast scanning of the systems, this value can be used to select ranges (e.g. 100-105) or to quickly find a specific system ID to debug.

    The \textit{unique system ID} (\texttt{UID}) can instead be any string and can be used to identify systems across different files. The only requirement for the \texttt{UID} is that two systems cannot have the same \texttt{UID} within the same file. It can be used, for example, to match the system described in the standard format with the analogous one in the original file of the specific BPS code output.
\end{fancyBox}

\begin{simpleBox}{Forward-compatibility: the Table of Values}{red}
To ensure forward-compatibility, the only requirement is that \textbf{all the columns described in Table \ref{tab:columns} are present in the Table of Values}. The labels are case-sensitive and must be written exactly as in this document. The order in which they appear is not fixed. Moreover, additional columns that are not described in Table \ref{tab:columns} and not codified in an extension are ignored. Additional labels are thus not forbidden, while not accounted for by the standard. In this way, future versions of the format that add new columns will be compatible with parsers that are compliant with older versions.
\end{simpleBox}

\subsubsection{Component types}
\label{sec:component_types}
A component type is a numerical label that describes the evolutionary state of one of the two components of the binary system. The description can be generic (e.g. star) or detailed (e.g. asymptotic giant branch star), and this is achieved by placing the labels in a tree-like structure, where parents are more generic than children and leaves are the most specific labels. The numerical value of the label mimics the structure of the tree. From the leftmost digit, indicating the most generic branch, to the rightmost digit which indicates the most specific leaf. For example, the numerical label for a first giant branch star is {\color{red} 1}{\color{teal} 2}{\color{blue}3}, where {\color{red}1} in the first digits indicates that it is a star, {\color{teal}2} in the second digit means that it has a hydrogen envelope and {\color{blue}3} that it is currently burning hydrogen in a shell.

At each level on the tree, the number 9 is reserved for 'other', and can be used in case more than 8 elements are needed at a particular level. The additional ones can be labelled 91, 92.. etc. Future extensions to the format will preferably follow the same labelling pattern.

The following numerical labels are supported by this version of \texttt{BinCodex}. Any label that is not in the following list shall be considered equivalent to -2 (unknown/missing).

\begin{itemize}
    \item 1 - Star: This category contains all objects that are undergoing some form of nuclear fusion in their interior and can be classified as stars
    \begin{itemize}
        \item 11 - Protostar: an object of stellar mass that has not reached the zero age main sequence.
        \item 12 - Star with a hydrogen envelope
        \begin{itemize}
            \item 121 - Main sequence (or core hydrogen burning)
            \item 122 - Hertzsprung gap
            \item 123 - First giant branch (or shell hydrogen burning)
            \item 124 - Core helium burning
            \item 125 - Asymptotic giant branch (or shell helium burning)
            \begin{itemize}
                \item 1251 - Early asymptotic giant branch
                \item 1252 - Thermally pulsing asymptotic giant branch
            \end{itemize}
        \end{itemize}
        \item 13 - Helium star: A star that is completely depleted of a hydrogen envelope
        \begin{itemize}
            \item 131 - Helium main sequence
            \item 132 - Helium Hertzsprung gap
            \item 133 - Helium first giant branch
        \end{itemize}
        \item 14 - Carbon star: A star that is completely depleted of both hydrogen and helium envelope
        \item 15 - Chemically homogeneous star: star that maintains a negligible internal chemical gradient due to efficient chemical mixing processes.
    \end{itemize}
    \item 2 - White dwarf
    \begin{itemize}
        \item 21 - Helium white dwarf
        \item 22 - Carbon-Oxygen white dwarf
        \item 23 - Oxygen-Neon white dwarf
    \end{itemize}
    \item 3 - Neutron star
    \item 4 - Black hole
    \item 5	- Planet
    \item 6 - Brown dwarf
    \item 7	- Thorne-\.Zytkow Object: is a possible result of the merger of a star and a neutron star. A star with an accreting neutron star in the core.
    \item 9 - Other: anything that does not fit in any previous category.
    \item -1 - Massless remnant: this object does not exist anymore. e.g. it has exploded or has been disrupted or has merged with the companion. This label serves as a placeholder.
    \item -2 - Unknown: the component type of the object is not specified. This information is missing.
\end{itemize}

\subsubsection{Events}
\label{sec:events}
A state event is a numerical label that describes what happened in the system that triggered the output of the current line ( or equivalently the output of an interesting state).
The numbering system is similar to the one used for the component type.

Hereafter, the character \textasteriskcentered~ stands for 0, 1, 2 or 3 with the meaning
\begin{itemize}
    \item 0 - not specified
    \item 1 - component 1
    \item 2 - component 2
    \item 3 - both components
\end{itemize}

The following numerical labels for events are supported by this version of \texttt{BinCodex}. Any label that is not in the following list shall be considered equivalent to -2 (unknown/missing).

\begin{itemize}
    \item 1\textasteriskcentered~- Component \textasteriskcentered~ changes type
    \item 2\textasteriskcentered~ - Component \textasteriskcentered~ goes supernova
    \begin{itemize}
        \item 2\textasteriskcentered1 - runaway thermonuclear explosion in degenerate matter (type Ia supernova). 231 for a double degenerate scenario, while in a single degenerate scenario \textasteriskcentered~ is the index of the accretor.
        \item 2\textasteriskcentered2 - Core collapse supernova
        \item 2\textasteriskcentered3 - Electron capture supernova
        \item 2\textasteriskcentered4 - Pair-instability supernova
        \item 2\textasteriskcentered5 - Pulsational pair-instability supernova
        \item 2\textasteriskcentered6 - Failed supernova (direct collapse into a black hole)
    \end{itemize}
    \item 3\textasteriskcentered~ - Component \textasteriskcentered~ overflows its Roche lobe
    \item 4\textasteriskcentered~ - Component \textasteriskcentered~ goes back into its Roche lobe (end of stable mass transfer or end of a contact phase)
    \item 5 - the surface of the two components touch.
    \begin{itemize}
        \item 51\textasteriskcentered~ - Component \textasteriskcentered~ engulfs the companion, triggering a common envelope (513 is the double common envelope).
        \item 52 - the two components merge.
        \item 53 - the system initiates a contact phase.
        \item 54 - the two components collide at periastron.
    \end{itemize}
    \item 8 - terminating condition reached, evolution is stopped. Only the last line of the evolution of a system can have an event with 8 as the first digit. If more than one condition applies at the final step, any of them can be provided. 

    Events starting with 8 should only be used to provide a reason for termination, and never to describe a physical change in the system: when a terminating condition is associated with an event, the format requires to first print the event and then repeat the line with a terminating condition. For instance, if only one object is left (code 84) because of a merger (code 52) the output will have first a line with code 52, and then another one with code 84.
    \begin{itemize}
        \item 81 - max time reached
        \item 82 - both components are compact remnants
        \item 83 - the binary system is dissociated
        \item 84 - only one object is left (e.g. due to a merger or because the companion has been disrupted)
        \item 85 - nothing left (both components are massless remnants)
        \item 88 - the evolution failed. This (and anything starting with 88) can be used as an error code.
        \item 89 - other: a terminating condition different from any previous one
    \end{itemize}
    \item 9 - other: any event that does not fit in any previous category.
    \item -1 - no notable events happened in this time step. This can be used when the output of the code is not based on events but is given at fixed time intervals.
    \item -2 - unknown: the event is not specified. This information is missing.
\end{itemize}

\section{Log file for system-specific parameters}
\label{sec:system-specific parameters}
Specific applications focused on parameter exploration could produce output files mixing binary systems having a different set of physical parameters or exploring combinations of physical assumptions. In this scenario, every system in a given file is in principle produced by a  set of different parameters. For example, a realistic sample of Galactic binaries will have a large number of systems at different initial metallicity. The conditions of each binary should be carefully traced for their comparison and physical interpretation, while the size of the dataset certainly impedes their description in comment lines, although allowed by the standard. 

As a scalable workaround, the present version of the standard format allows the existence of a log file, having the same name as the data file and the ".log" extension, in which the ID and UID are associated with all the required parameters. 

The present document does not standardize the content of the log file, which is left free to the specific needs of BPS code implementors.

As a usage example, a log file could be adopted to store the initial conditions generating a binary system set. Table \ref{tab:ICTab} provides an example of initial conditions storable in a .log file.

\begin{table}
\label{tab:ICTab}
\begin{tabular}{|c|l|c|c|}
\hline
Label & Description & Unit & Variable Type\\
\hline
\texttt{ID}            & progressive system ID   &             & long   \\
\texttt{UID}           & unique system ID        &             & string \\
\texttt{semiMajor}     & semi-major axis         & $R_\odot$   & double \\
\texttt{eccentricity}  & orbital eccentricity    &             & double \\
\texttt{type1}         & type of object 1        &             & string    \\
\texttt{mass1}         & mass 1                  & $M_\odot$   & double \\
\texttt{Z1}            & absolute metallicity 1  &             & double \\
\texttt{type2}         & type of object 2        &             & string    \\
\texttt{mass2}         & mass 2                  & $M_\odot$   & double \\
\texttt{Z2}            & absolute metallicity 2  &             & double \\
\hline
\end{tabular}
\caption{An example of columns for the initial condition file}
\end{table}

\section{User or BPS-specific Extensions}
\label{sec:extensions}

The present version of \texttt{BinCodex} is designed to be compatible with user-specific or BPS-specific extensions. An extension could be required, for example, when a combination of values does not conform to either T1 or T2 or when it is convenient to mix a certain common format tier (e.g. T0) with BPS-specific outputs. In this case, the data file will be declared as conforming to T0, extended with a documented set of parameters. When a code extension becomes adopted by all BPS codes, the extension implementors could consider asking for its inclusion at the T0 tier of the common format and the extension becomes part of a new format release. A widely used extension could also be included in either T1 or T2 depending on its popularity across BPS codes.

The following classes of extensions are allowed in the present version:
\begin{itemize}
    \item addition of fields in the Header,
    \item addition of new columns in the Table of Values,
    \item definition of new component types and event types
\end{itemize}

\subsection{Defining an Extension}
Users or BPS code implementors can define their own extensions of the present format by writing a document containing
\begin{enumerate}
    \item the name of the extension,
    \item the unique labels of the new columns/header fields, their variable type and a short description of their meaning.
    \item the unique numerical labels of the new component types/event types and a short description of their meaning.
\end{enumerate}

This document will be shared as an additional file and to anyone who needs to use the dataset implementing the described extension.
The common format does not specify the extension types, values or events but requires them to be written as columns/values successive to the last value of the adopted tier. File parsers will safely work with columns codified by the tier format (e.g. T1) and should be modified to understand extension columns or labels.

\subsection{Using an Extension}
To use an extension
\begin{enumerate}
    \item write the name of the extension in the field \texttt{cofExtension} in the Header. When no extension is used the field is left empty.
    \item add the labels of the new columns/header fields to the Header.
    \item fill those columns/header fields with the newly defined quantities.
    \item use the new component types/event types labels in the appropriate column, when necessary.
\end{enumerate}

%\section*{Changes from the previous version (0.2)}
%\begin{itemize}
%    \item Changed nomenclature from Levels (L0, L1...) to Tiers (T0, T1...), to avoid confusion with ESA missions.
%    \item Separated the T1 field \texttt{EnvBindEn} into two fields \texttt{envBindEn1} and \texttt{envBindEn2} for the two stars.
%    \item Chosen the name \texttt{BinCodex} for the format.
%    \item Specified requirement that event codes of type 8 should only be used to inform about terminating conditions and not as descriptions of physical events.
%\end{itemize}

\newpage

\LTcapwidth=\textwidth
\begin{longtable}{|l|>{\itshape\small}p{0.6\textwidth}|}
 \hline
 Name & {\normalfont\normalsize Affiliation}\\
 \hline
 \endfirsthead

 \hline
 \multicolumn{2}{|c|}{Continuation of Table \ref{tab:my_label}}\\
 \hline
 Name & {\normalfont\normalsize Affiliation}\\
 \hline
 \endhead
\hline
\endfoot
\hline
\multicolumn{2}{c}{}\\
\caption{\label{tab:my_label}The LISA Synthetic UCB Catalogue Group members in alphabetical order}\\
\endlastfoot

        Agrawal, Poojan \orcid{0000-0002-1135-984X}
        & 
        Department of Physics and Astronomy, The University of North Carolina at Chapel Hill, Chapel Hill, NC 27599, USA \\
        \hline
        Arca Sedda, Manuel \orcid{} &
        Gran Sasso Science Institute, V.le F. Crispi 67100 L'Aquila, Italy\newline
        INFN – Laboratori Nazionali del Gran Sasso, 67100 L’Aquila, (AQ), Italy\newline
        INAF – Osservatorio Astronomico d’Abruzzo, Via M. Maggini snc, 64100 Teramo, Italy        
        \\
        \hline
        Artale, M. Celeste \orcid{0000-0003-0570-785X} &
        Instituto de Astrof\'isica, Facultad de Ciencias Exactas, Universidad Andres Bello, Fernandez Concha 700, Santiago, Chile\\
        \hline
        Bobrick, Alexey \orcid{0000-0002-4674-0704} & Physics Department \& Technion - Israel Institute of Technology, Haifa, 3200002, Israel  \\
        \hline
        Breivik, Katelyn \orcid{0000-0001-5228-6598} & Department of Physics \& McWilliams Center for Cosmology,\newline Carnegie Mellon University, \newline 5000 Forbes Avenue, Pittsburgh, PA 15213, USA \\
        \hline
         Chakraborty, Srija \orcid{0000-0002-4906-2670} & Department of Physics, Scuola Normale Superiore, \newline Piazza dei Cavalieri 7, 56126 Pisa PI, Italy\\
        \hline
        \c{C}al{\i}\c{s}kan, Mesut \orcid{0000-0002-9231-1505} & Department of Physics \& Astronomy, Johns Hopkins University,
        3400 North Charles Street, Baltimore, MD 21218, USA \\
        \hline
        Chen, Xuefei \orcid{0000-0001-5284-8001}
        & Yunnan Observatories, Chinese Academy of Sciences, Kunming, 650011, People's Republic of China\newline
        Key Laboratory for the Structure and Evolution of Celestial Objects, Chinese Academy of Science, People's Republic of China\newline
        International Centre of Supernovae, Yunnan Key Laboratory, Kunming, 650216, People's Republic of China\\
        \hline
        Eldridge, Jan J. \orcid{0000-0002-1722-6343}
        & Department of Physics, University of Auckland, Private Bag 92019, Auckland, New Zealand \\
        \hline
        Escobar, Gast\'on J. \orcid{0000-0002-4007-7585}
        & Dipartimento di Fisica e Astronomia, Universit{\`a} di Padova, \newline Vicolo dell'Osservatorio 3, 35122, Padova, Italy  \\
        \hline
        Gossage, Seth \orcid{0000-0001-6692-6410} & Center for Interdisciplinary Exploration and Research in Astrophysics (CIERA), Northwestern University, \newline 1800 Sherman Ave, Evanston, IL 60201, USA \\
        \hline
        Graziani Luca \orcid{0000-0002-9231-1505} & Dipartimento di Fisica, Sapienza, Universit{\`a} di Roma, \newline Piazzale Aldo Moro 5, 00185, Roma, Italy \\
        \hline
        Han, Zhanwen \orcid{0000-0001-9204-7778}
        & Yunnan Observatories, Chinese Academy of Sciences, Kunming, 650011, People's Republic of China\newline
        Key Laboratory for the Structure and Evolution of Celestial Objects, Chinese Academy of Science, People's Republic of China\newline
        International Centre of Supernovae, Yunnan Key Laboratory, Kunming, 650216, People's Republic of China\\
        \hline
        Iorio Giuliano \orcid{0000-0003-0293-503X} & Dipartimento di Fisica e Astronomia, Universit{\`a} di Padova, \newline Vicolo dell'Osservatorio 3, 35122, Padova, Italy \\
        \hline
        Korb Erika \orcid{0009-0007-5949-9757} & Dipartimento di Fisica e Astronomia, Universit{\`a} di Padova, \newline Vicolo dell'Osservatorio 3, 35122, Padova, Italy \\
        \hline
        Korol Valeriya \orcid{0000-0002-6725-5935} &  Max-Planck-Institut f{\"u}r Astrophysik, \newline Karl-Schwarzschild-Straße 1, 85741 Garching, Germany \\
        \hline
        Kruckow, Matthias~U. \orcid{0000-0001-9331-0400} & D\'{e}partement d'Astronomie, Universit\'{e} de Gen\`{e}ve, Chemin Pegasi 51, CH-1290 Versoix, Switzerland\newline Gravitational Wave Science Center (GWSC), Universit\'{e} de Gen\`{e}ve, CH-1211 Geneva, Switzerland\\
        \hline
         Larson, Shane\ L. \orcid{0000-0001-7559-3902} 
         & Center for Interdisciplinary Exploration \& Research in Astrophysics, and Department of Physics \& Astronomy, Northwestern University, Evanston, IL 60201, USA  \\
        \hline
        Laurentiu, Caramete \orcid{0000-0002-3571-3145}
        & Cosmology and Astroparticle Physics Laboratory, Institute of Space Science, Atomistilor Street 409, 077125, Magurele, Romania\\
        \hline
        Li, Zhenwei \orcid{0000-0002-1421-4427}
        & Yunnan Observatories, Chinese Academy of Sciences, Kunming, 650011, People's Republic of China\newline
        Key Laboratory for the Structure and Evolution of Celestial Objects, Chinese Academy of Science, People's Republic of China\newline
        International Centre of Supernovae, Yunnan Key Laboratory, Kunming, 650216, People's Republic of China\\
        \hline
        Mandel, Ilya \orcid{0000-0002-6134-8946} &
        School of Physics and Astronomy, Monash University, Clayton, VIC 3800, Australia\newline
        The ARC Centre of Excellence for Gravitational Wave Discovery -- OzGrav, Australia\\
        \hline
        Menu, Jonathan \orcid{0000-0001-6905-1840} & 
        Institute for Theoretical Physics, KU Leuven,\newline Celestijnenlaan 200D, 3001 Leuven, Belgium\\
        \hline
        Nelemans, Gijs \orcid{0000-0002-0752-2974} & Department of Astrophysics/IMAPP, Radboud University,  P.O. Box 9010, 6500 GL, Nijmegen, the Netherlands\newline
        Institute of Astronomy, KU Leuven, Celestijnenlaan 200D, 3001 Leuven, Belgium\newline
        SRON, Netherlands Institute for Space Research, Niels Bohrweg 4, 2333 CA Leiden, The Netherlands\\
        \hline
         Rodriguez-Segovia, Nicolas \orcid{0000-0002-0125-1472} & 
         School of Science, University of New South Wales, Australian Defence Force Academy, Canberra, ACT 2600, Australia\\
        \hline
         Ruiter, Ashley J. \orcid{0000-0002-4794-6835} & 
         School of Science, University of New South Wales, Australian Defence Force Academy, Canberra, ACT 2600, Australia\\ 
         \hline
        Tang, Petra \orcid{0000-0001-5743-9733} & Department of Physics, University of Auckland, Private Bag 92019, Auckland, New Zealand \\
        \hline
         Toonen, Silvia \orcid{0000-0002-2998-7940} & 
         The Anton Pannekoek Institute for Astronomy, University of Amsterdam (UvA), Science Park 904, 1098 XH Amsterdam, Netherland. \\
        \hline
        Sgalletta, Cecilia \orcid{0009-0003-7951-4820} & SISSA, via Bonomea 365, I–34136 Trieste, Italy\\
        \hline
        Strokov, Vladimir \orcid{0000-0002-6555-8211} & Department of Physics \& Astronomy, Johns Hopkins University,
        3400 North Charles Street, Baltimore, MD 21218, USA \\
        \hline
        Sun, Meng \orcid{0000-0001-9037-6180} & Center for Interdisciplinary Exploration and Research in Astrophysics (CIERA), Northwestern University, 1800 Sherman Ave,
Evanston, IL 60201, USA \\
        \hline 
        Valli Ruggero \orcid{0000-0003-3456-3349} &  Max-Planck-Institut f{\"u}r Astrophysik, \newline Karl-Schwarzschild-Straße 1, 85741 Garching, Germany \\
        \hline
        Willcox, Reinhold \orcid{0000-0003-0674-9453} &  
        Institute of Astronomy, KU Leuven, Celestijnenlaan 200D, 3001 Leuven, Belgium\newline
        School of Physics and Astronomy, Monash University, Clayton, VIC 3800, Australia \\
 \end{longtable}

\end{document}